\documentclass[11pt]{article}
\pdfoutput=1
\usepackage{amsmath,epsf,amssymb,latexsym,array,color}

\usepackage{graphicx,wasysym}
\usepackage{cancel}
\usepackage{stmaryrd}
\usepackage{hyperref}
\usepackage[compat=1.1.0]{tikz-feynman}
\usepackage{cite}

\def\be{\begin{equation}}   \def\ee{\end{equation}}
\def\ba{\begin{eqnarray}}   \def\ea{\end{eqnarray}}
\def\nn{\nonumber}   \def\lan{\langle} \def\ran{\rangle}
	\def\ov{\over}	\def\ra{\rightarrow}

\def\ms{ \overline {\rm MS}}
\def\lms{ \Lambda_{\ms}}	
	
\def\ran{\rangle}	\def\lan{\langle}

\newcommand{\bbox}{\lower.2ex\hbox{$\Box$}}
\csname @addtoreset\endcsname{equation}{section}
\usepackage{mciteplus}
\usepackage{color}

\def\be{\begin{equation}}
\def\ee{\end{equation}}

\usepackage{enumitem} % 프리앰블에 추가

\begin{document}

\begin{titlepage}

\hskip 1.5cm

\begin{center}
{\huge\bf{The QCD Scale Parameter from the Photon Structure Function}}
\vskip 0.8cm  
{\bf \large Hun Jang\footnote{hun.jang@nyu.edu}$^{a,b}$, Eun Bok$^{c}$,  Hyeunwoo Kim$^{d}$, \\ Byeongjun Yoon$^{d}$, and Sun Myong Kim$^{d}$\footnote{minntenn@yonsei.ac.kr}}  
\vskip 0.75cm
{\em $^{a}$Center for Quantum Spacetime (CQUeST), Sogang University, Seoul 04107, Republic of Korea\\
  $^{b}$Yukawa Institute for Theoretical Physics (YITP), Kyoto University, Kyoto 606-8502, Japan  \\
 $^{c}$Institute of Advanced Machines and Design, Seoul National University,\\ 1 Gwanak-ro, Gwanak-gu, Seoul 08826, Republic of Korea\\
 $^{d}$Department of Physics and Engineering Physics, Yonsei University,\\ Wonju, Kangwon-Do 26493, Republic of Korea}
\vspace{12pt}

\end{center}

\begin{abstract}
The photon structure function has been a solid platform for testing the strong interaction
along with the nucleon structure function.
The strong interaction has the property that it is perturbatively calculable at high
energy but becomes non-perturbative at low energy.
This nature makes QCD difficult to handle theoretically regarding the factorization of these two regions.
The fundamental dimensional parameter, so called the QCD scale parameter $\lms$,
is a key player in factorizing these two energy regions.
In this work, we extract the QCD scale parameter from the photon structure function
by separating the perturbative and non-perturbative QCD contributions.
To achieve this, we use the vector dominance model for the non-perturbative energy region
of the photon structure function.
\end{abstract}

\vskip 1 cm
\vspace{24pt}
\end{titlepage}
\tableofcontents

%%%%%%%%%%%%%%%%%%%%%%%%%%%%%%%%%%%%%%%%%%%%%%%%%%%%%%%%%%%%%%%%%%
\section{Introduction}

The discovery of asymptotic freedom \cite{asyfree1,asyfree2} in the strong interaction has made
quantum chromodynamics (QCD) successful in describing nucleon deep inelastic scattering (DIS),
such as $ep\ra eX$  at HERA, $pp\ra X$ or $p\bar p\ra X$ at Fermilab and CERN, involving various hadrons $X$.
Consequently, QCD has become the most powerful perturbation theory
for strongly interacting high energy phenomena.
However, at energy scales corresponding to hadron formation (approximately a few GeV or less),
the QCD perturbation theory ceases to work due to the nature of the strong interaction.
In this regime, perturbative QCD (PQCD) can no longer be applied alone, 
and non-perturbative QCD (NP) should also be carefully taken into account.

There exists a fundamental dimensional constant, $\lms$,
which provides the criterion for separating two energy regions.
This constant is defined in the modified minimal subtraction renormalization scheme expressed as $\ms$.
Because this constant lies at the boundary of these energy regions,
determining its value accurately-both theoretically and experimentally-remains a significant challenge.
Consequently, the precise determination of the constant is a long-standing problem.
While it has primarily been explored through nucleon deep inelastic scattering,
Ibes and Walsh \cite{IW} proposed a method to extract $\lms$ from the photon
structure function (PSF).
Their approach is based on the Bjorken separation \cite{Bjorken} of NP components from the PSF.

Although both nucleon deep inelastic scattering (DIS)
and two photon scattering via hadrons involve processes with $\lms$,
the two photon process has the advantage of having no compositeness in its initial particles compared to DIS.
Furthermore, two photon scattering exhibits both perturbative and non-perturbative QCD features
at relatively high $Q^2$, with the probe photon momentum $q$ with $q^2=-Q^2$
and the target photon momentum $p$ with $p^2=-P^2$.
This is the reason why we are interested in exploring QCD in PSF.
This may make the two photon scattering a unique
process in which we can study both of these parts of PSF with the kinematic variables.
Uematsu and Walsh proposed the perturbative part of the struncture function \cite{UW}
while authors in \cite{Newman,NA3,CD} proposed the non-perturbative hadronic part.

The operator product expansion (OPE) and
the renormalization group equation (RGE) have been used in the high order calculation of PSF.
They are particularly useful for the perturbative PSF, which can be expressed in terms of moments, $f(n,Q^2)$ \cite{UW,BB1,BB2,KW}.
Working with these moments expressed in $n$ is easier than treating the structure function directly in $x$
in higher order calculations.
This is the reason why we use the moments of the structure function and
subsequently invert them to $x$-space to obtain the structure function $F(x,Q^2)$.

The basic idea of this work is to examine the behavior of the total PSF (including both PQCD and NP parts) as a function of 
$Q^2$ and $P^2$ with the goal to distinguish these two parts theoretically and experimentally.
First, we  describe the model for the NP contribution how it is related to the real PSF and the virtual PSF
with the $Q^2$ and $P^2$ dependence.
We then present our results how the $P^2$ dependence can be employed to the virtual photon structure function. 
We will follow the convention and the definitions of \cite{UW} throughout unless otherwise stated.

In Sec. 2, we discuss PSF, $\lms$, their relation, and how $\lms$ can be extracted from PSF in general.
In Sec. 3, vector meson dominance (VMD) is introduced
to separate the perturbative (PQCD) part and non-perturbative (NPQCD) contributions to the PSF.  
We discuss the numerical inversion of moments and present the extracted values of $\lms$ from the PSF in Sec. 4.
Then, we discuss the work to be done experimentally and theoretically in the future.
We conclude our work in Sec. 5.

%%%%%%%%%%%%%%%%%%%%%%%%%%%%%%%%%%%%%%%%%%%%%%%%%%%%%%%%%%%%%%%%%%
\section{Photon Structure Function and the Scale Parameter $\Lambda_{\overline{\textrm{MS}}}$}
%%%%%%%%%%%%%%%%%%%%%%%%%%%%%%%%%%%%%%%%%%%%%%%%%%%%%%%%%%%%%%%%%%

There have been numerous
successful applications of QCD to hadronic processes since the discovery of asymptotic freedom.
In particular, the perturbative method in QCD has been very fruitful in
explaining hadronic processes in high energy physics.
On the other hand, we are still attempting to understand many of hadronic processes that
contain relatively large non-perturbative contributions, which we do not know how to calculate,
although some results have been derived from the lattice gauge theory.

In deep inelastic scattering, a very energetic probe particle transfers the large momentum ($q$) to the target particle.
This momentum transfer is much larger than the fundamental constant, so-called QCD scale parameter ($-q^2=Q^2 \gg \lms^2$),
so that we are able to apply the perturbative QCD to the process.
%From the nature of $\lms$, presumably the energy scale of starting color confinement, however, it is necessary to consider
However, because $\lms$ represents the energy scale where  color confinement begins, it is necessary to consider
the structure function at moderately low value of $Q^2$ where we
always have incalculable non-perturbative terms which make it difficult to further analysis.
The best thing we can do at present is to separate the perturbative and non-perturbative parts.
Although the job is not so trivial, we assume that
the present calculations give good approximation of the perturbative component of structure function.

The scattering channels $ep\ra eX$ or $pp\ra X$ or $p\bar p\ra X$ ($X$: hadron(s)) are widely investigated for hadronic processes.
The proton, $p$, consists of quarks and gluons which interact via the strong interaction. 
This makes it difficult to deal with these hadronic processes.
On the other hand, another hadronic process, $ee\ra eeX$, is much simpler to analyze since the initial particles are only leptons.
This process can be analyzed further as $e^+ e^- \ra e^+ e^- \gamma\gamma^*\ra e^+ e^- X$.
The typical hadronic analysis for this scattering involves two photons as shown in Fig. 1.
The substructure of this process contains two photon process,
$\gamma^*\gamma\ra X$, or $\gamma^*\gamma^*\ra X$.
Where $\gamma$ and $\gamma^*$ are real and virtual photons respectively,
and $X$ represents hadrons mostly produced via quarks ($q\bar{q}$ in Fig. 1).
One photon ($\gamma^*$) is considered a probe photon and the other photon ($\gamma$ or $\gamma^*$)
as a target photon in the photon-photon scattering.\\
%%%%%%%%
\begin{figure}[h]                             
%	\centering
\begin{center}
\begin{tikzpicture}[scale=1.6, transform shape, line width=1.3pt]
  \begin{feynman}
    %--- 전자선 (왼쪽 in/out) ---
    \vertex (e1in)  at (-2,  1.2) {\(e\)};
    \vertex (e1v)   at (-0.5, 1.2);
    \vertex (e1out) at ( 0.8,  1.8) {\(e\)};
    \vertex (e2in)  at (-2, -1.2) {\(e\)};
    \vertex (e2v)   at (-0.5,-1.2);
    \vertex (e2out) at ( 0.8, -1.8) {\(e\)};

    %--- 쿼크선 (오른쪽 q, qbar + 내부 전파선) ---
    \vertex (vtop)    at (1.5,  0.6);
    \vertex (vbot)    at (1.5, -0.6);
    \vertex (q)       at (3.2,  0.6) {\(q\)};
    \vertex (qbar)    at (3.2, -0.6) {\(\bar q\)};

    \diagram*{
      % 전자선
      (e1in)  -- [fermion] (e1v)  -- [fermion] (e1out);
      (e2in)  -- [fermion] (e2v)  -- [fermion] (e2out);

      % 광자선 (t-channel)
      (e1v) -- [photon, edge label=\(\gamma^{*}\)] (vtop);
      (e2v) -- [photon, edge label'=\(\gamma(\gamma^{*})\)] (vbot);

      % 쿼크선
      (vbot) -- [fermion] (vtop);       % 내부 전파선 (화살표 위쪽)
      (vtop) -- [fermion] (q);          % q
      (qbar) -- [fermion] (vbot);       % \bar q (반입자라 반대방향)
    };
  \end{feynman}
\end{tikzpicture}
\end{center}
\caption{$ee \rightarrow eeX$ via quarks to produce the hadron $X (=q\bar{q})$. The substructure of this process contains the two photon process $\gamma\gamma\ra q\bar{q}$ at tree level eventually $q\bar{q}\ra X$.}	\label{figure1}	
\end{figure}
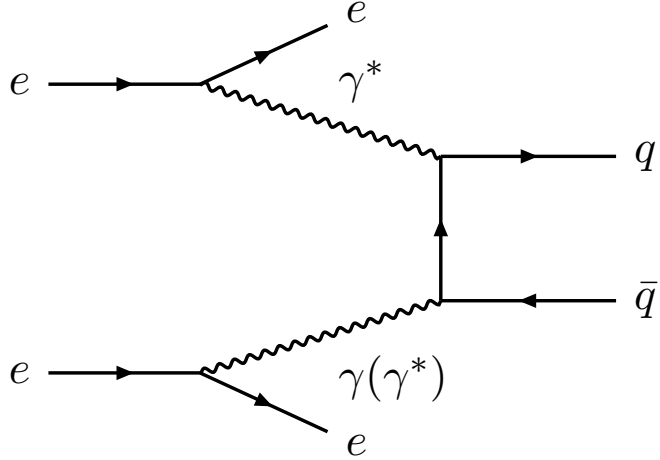

%============ 
\noindent {\bf  Scattering Amplitude for Photon Photon interaction.}
Like the proton (or nucleon) structure function in the deep inelastic scattering such as $ep\ra eX$,
the photon can also be considered to have a structure.
In a two-photon process, the probe photon is highly virtual, with a momentum $q$ satisfying $q^2=-Q^2<0$,
while the target photon can be either real with the momentum $p$ ($p^2=0$) or virtual
with the momentum $p$ ($p^2=-P^2<0$).
We call the structures for these target photons the real photon structure function (rPSF)
and the virtual photon structure function (vPSF), respectively.
The vPSF reduces to the rPSF when the target momentum is set to $p^2=0$.

The structure function of the photon was found in parton level by Walsh and Zerwas \cite{WZ}, 
in the leading order in QCD by Witten \cite{W},
and in the next-to-leading order in QCD by Bardeen and Buras \cite{BB1,BB2}.
The more general analysis of the vPSF can be found in \cite{UW}.

%%%%%%%%
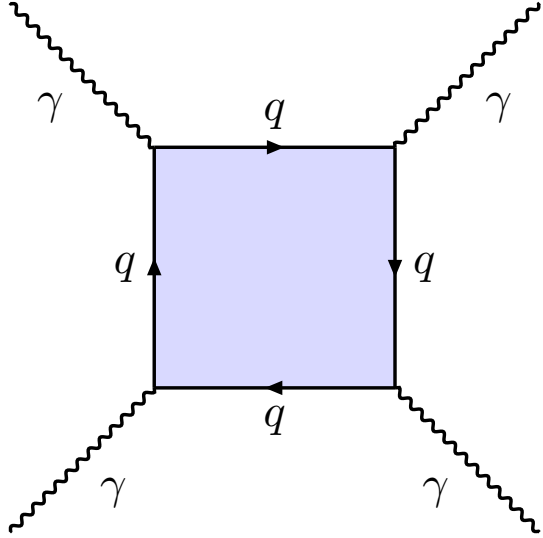
\begin{figure}[h]                             
%	\centering
\begin{center}
\begin{tikzpicture}[scale=1.6, transform shape, line width=1.3pt]

  % 파란 박스 (배경)
  \fill[blue!15] (-1,-1) rectangle (1,1);

  \begin{feynman}
    % 박스 꼭짓점 (쿼크 루프)
    \vertex (tL) at (-1,  1);
    \vertex (tR) at ( 1,  1);
    \vertex (bR) at ( 1, -1);
    \vertex (bL) at (-1, -1);

    % 외부 photon 끝점 (대각 방향)
    \vertex (gTL) at (-2.2,  2.2);
    \vertex (gTR) at ( 2.2,  2.2);
    \vertex (gBR) at ( 2.2, -2.2);
    \vertex (gBL) at (-2.2, -2.2);

    \diagram*{
      % q 루프 (시계 방향)
      (tL) -- [fermion, edge label=\(q\)] (tR)
           -- [fermion, edge label=\(q\)] (bR)
           -- [fermion, edge label=\(q\)] (bL)
           -- [fermion, edge label=\(q\)] (tL);

      % γ 외선 (꼭짓점에서 나가는 photon)
      (gTL) -- [photon, edge label'=\(\gamma\)] (tL);
      (gTR) -- [photon, edge label=\(\gamma\)]  (tR);
      (gBR) -- [photon, edge label=\(\gamma\)]  (bR);
      (gBL) -- [photon, edge label'=\(\gamma\)] (bL);
    };
  \end{feynman}
\end{tikzpicture}
\end{center}
\caption{Box Diagram for two photon process. The scattering amplitude for this box diagram has no internal structure at tree level.
However, quantum fluctuations due to QCD interactions produce a very complicated inner structure inside the box,
namely, complex gluon connections within the quark lines that form the box.
}
\label{figure2}	
\end{figure}

According to Uematsu and Walsh \cite{UW}, the scattering amplitude for two virtual photons becomes
\be
T_{\mu\nu\alpha\beta}=i\int d^4x d^4y d^4z e^{iqx}e^{ip(y-z)}
\lan0|T(J_\mu(x)J_\nu(0)J_\alpha(y)J_\beta(z)|0\ran
\ee
and the structure tensor is the absorptive part of the amplitude in Fig. 2, 
\ba
W_{\mu\nu\alpha\beta}(p,q)&=&{1\over\pi}{\rm Im}T_{\mu\nu\alpha\beta}\nn\cr
 &=&{1\over2}\int d^4x d^4y d^4z e^{iqx}e^{ip(y-z)}
\lan0|T^*(J_\mu(x)J_\alpha(y))T(J_\nu(0)J_\beta(z))|0\ran.
\ea
The structure tensor has eight independent structure functions \cite{BM,CT}.
After a spin average for the target photon,
these eight independent structure functions are reduced to two independent structure functions: $F_2(x,Q^2,P^2)$ and $F_L(x,Q^2,P^2)$.
We consider only $F_2$, as $F_L$ is small compared to $F_2$.
Where a virtual photon target whose invariant mass square,
$ - P^2$, can vary from zero to an appreciable $P^2 \ll Q^2$.\\

\noindent {\bf Operator Product Expansion and Renormalization Group Equation.}
It is not an easy task to obtain structure functions in higher order in terms of the Bjorken variable $x$.
Witten used the operator product expansion (OPE) to obtain PSF in the leading order \cite{W}.
Later, Bardeen and Buras \cite{BB1,BB2} and Uematsu and Walsh \cite{UW} obtained
the rPSF and the vPSF in the next-to-leading order, respectively.
For that purpose, they introduce the Mellin transform to convert the structure functions of variable $x$ into the structure functions of moments $n$,
\ba
F_2^\gamma(n,Q^2,P^2)&=&
\int_0^1 dx x^{n-2} F_2^\gamma(x,Q^2,P^2)\nn\\
&=&\sum_{i=\psi,G,NS}C_n^i(Q^2/\mu^2,g(\mu),\alpha)\lan\gamma(p)|O_n^i(\mu)|\gamma(p)\ran\nn\\
&& +~C_n^\gamma(Q^2/\mu^2,g(\mu),\alpha)\lan\gamma(p)|O_n^\gamma(\mu)|\gamma(p)\ran.
\ea
Where $\psi,~G,~NS,~\gamma$ stand for the singlet quark, gluon, non-singlet quarks, and photon, respectively.
$O_n^i$ and $O_n^\gamma$ are the operators, and 
$C_n^i(Q^2/\mu^2,g(\mu),\alpha)$ and $C_n^\gamma(Q^2/\mu^2,g(\mu),\alpha)$ are their corresponding coefficients.
Here, $g$ and $\alpha$ represent the strong and electromagnetic coupling constants, respectively.
In this calculation, we utilize several renormalization group equations
in terms of the energy scale variable $\mu$ \cite{Muta}.
The normalization point $\mu$ is chosen to be $\mu^2=P^2$.
The quark masses are neglected for sufficiently large $P^2$ and $Q^2$ \cite{HR:1979}.
Finally, the structure functions in $x$-space are obtained by numerically performing the inverse Mellin transform of the functions of $n$.\\

%============ 
\noindent {\bf The QCD scale parameter $\Lambda_{\overline{\textrm{MS}}}$.}
The strong coupling constant $\alpha_S$ is not actually a constant
but runs with the energy scale $\mu$, denoted as $\alpha_S(\mu)$.
The QCD (scale) parameter $\Lambda$ characterizes the strong coupling
so as to separate the perturbative and the non-perturbative regions.
In \cite{BBDM}, this parameter was redefined by multiplying it by a constant $\lms$.
 $\lms$ depends on the renormalization scheme, the order of the perturbation, and the number of active quark flavors, etc.
Determining this parameter has been a long-standing problem in QCD.
The constant has been extracted mostly from the $ep$ deep inelastic scattering.
Currently, the value of  $\lms$ is estimated to be in the range of 200 MeV to 400 MeV.
%which has been improved recently around 200 MeV.
%A few efforts to determine the value of $\lms$ can be found in \cite{lambda}.
For example, the value obtained in \cite{lambda1,lambda2} for four quark flavors is $\lms^{(4)}=205\pm 22(stat.)\pm 60(syst.)$.

However, deep inelastic scattering involving nucleons suffers from troublesome initial and final hadronic states.
On the other hand, two photon process from $ee$ or $ee^+$ scattering has simple initial leptonic states.
Ibes and Walsh \cite{IW} claimed that once all the available theoretical information is explored, 
uncertainties involved in determining the constant from experiment can be resolved,
rather than merely settling for theoretical difficulties.
Therefore, it is advantageous to extract the scale constant in PSF of this process theoretically, supported by experimental data.
We use  the most popular QCD renormalization scheme, $\ms$ (modified Minimal Subtraction) scheme \cite{BBDM, BB1, BB2},
for the calculation of PSF in this process.
We can determine the scale parameter under this scheme
with minimal uncertainty compared to other methods currently available.\\

%============ 
\noindent {\bf Separation of PSF into PQCD and NP parts.}
The QCD parameter separates the perturbative and non-perturbative parts of PSF.
The advantage of this decomposition is that the PQCD part is calculable, and the
non-perturbative part is experimentally determinable.
Unlike proton structure function, PSF has a well-established phenomenological model for the NP component.
Since vector mesons ($\rho, \omega, \phi$) have the same quantum numbers as the photon,
it isnatural to adopt the traditional method of treating the vector meson structure function as the NP part of the PSF.\\

%============ 
\noindent {\bf How to extract $\lms$.}
Now, how do we extract $\lms$ from PSF?
Even though we do not know how $\lms$ is involved in the hadronic part of the structure function, we do know its role in the PQCD part.
Therefore, isolating the PQCD part from the total PSF is very important.
We will summarize the method and the process to extract $\lms$ from the PSF in \cite{Kim}.
It is better to use more general vPSF rather than the rPSF, which can be obtained by taking $P^2=0$ from the vPSF.
The first step is to separate
the structure functions into a QCD-calculable perturbative piece and an incalculable non-perturbative one.
The non-perturbative part will be modeled by a vector meson, which can be extracted from experiment.
Separation of the NP part and the PQCD part in the PSF is not only useful but also essential in extracting $\lms$,
\be
F^\gamma(x,Q^2,P^2)=F^\gamma_{PQCD}(x,Q^2,P^2)+F^\gamma_{NP}(x,Q^2,P^2).
\label{tpsf}
\ee
The Mellin transform of Eq. (\ref{tpsf}) produces the moments of PSF depending on $n$,
\ba
F^\gamma(n,Q^2,P^2)&=&\int_0^1 x^{n-2}[F^\gamma_{PQCD}(x,Q^2,P^2)+F^\gamma_{NP}(x,Q^2,P^2)],\nn\\
&=&F^\gamma_{PQCD}(n,Q^2,P^2)+F^\gamma_{NP}(n,Q^2,P^2).
\ea
Where $F^\gamma_{NP}$ can be expressed as a function of $P^2$ in the kinematic region of $\lms^2 \ll P^2 \ll Q^2 $.
%\be
%F^\gamma_{NP}(n,Q^2,P^2)=\sum D^k_n(Q^2)\bigg({\lms\over P}\bigg)^2.
%\ee
%The coefficients $D^k_n(Q^2)$ is expected to be small at large $n$.
%Therefore, the higher order terms of $P^2$ are small for $\lms^2 \ll P^2$  
%and the terms in orders of $P^2/Q^2$ have been neglected.
%We, also, know from the experiment \cite{NPexp} that this NP part of PSF has soft $P^2$ dependence in the range of $\rm 0.5
%~GeV^2 - 2~GeV^2$ \cite{PWZ}.
%These properties of the NP part make us possible to extract $\lms$.
%
%When we choose $P^2$ large enough, the experimental value of PSF is close to the perturbative part of PSF,
%$ F^\gamma_{exp} \approx F^\gamma_{PQCD} $ at a fixed $Q^2$ ($\gg P^2$).
%In that case, we have very weak $\lms^2$ dependence in $F^\gamma_{PQCD}$.
%For example, $\rln Q^2/(P^2 + \lms^2) \approx \rln Q^2/P^2$ so that we can not
%determine $\lms$ with a reasonable accuracy.
%(Since the rPSF can be obtained from the vPSF by the formal replacement of $P^2$ to $\lms^2$, 
%we replace $P^2$ to $P^2+\lms^2$ in the denominator of the logarithm $\rln (Q^2/P^2)$ to avoid the uncertainty.)
%To get an reasonably accurate $\lms$ we need to go to $P^2=0$ region, which corresponds to the rPSF case.
We know $F^\gamma$ from the experiment \cite{NPexp}.
% and the NP part of PSF has soft $P^2$ dependence in the range of $\rm 0.5~GeV^2 - 2~GeV^2$ \cite{PWZ}.
Then, $F^\gamma_{NP}$ can be written as
\be
F^\gamma_{NP}(x,Q^2,P^2,\lms^2) = F^{\gamma}_{exp}(x,Q^2,P^2) - F^\gamma_{PQCD}(x,Q^2,P^2,\lms^2).
\label{Fnpexp}
\ee

Using experimental data for the PSF at various $P^2$ values for a fixed $Q^2$,
we determine $F^\gamma_{NP}$ via Eq. (\ref{Fnpexp}) at the experimental kinematics.
According to vector meson dominance, $F^\gamma_{NP}$ is related to the $\rho$ meson mass ($m_\rho\simeq 0.7$ GeV)
and varies smoothly over the range 0.5 GeV$^2$ - 2 GeV$^2$ \cite{PWZ1,PWZ2,BW:1987}.
Consequently, we can extrapolate $F^\gamma_{NP}$ at $P^2=0$ using experimental data.
This extrapolated value is independent of the renormalization scheme employed
in the perturbative contribution, $F^\gamma_{PQCD}$.
Therefore, scheme-dependent uncertainty is solely to the $F^\gamma_{PQCD}$ term.
We carry out the same experiment for the rPSF (corresponding to vPSF with $P^2=0$).
Once we measure the rPSF at the same fixed value of $Q^2$, we can obtain the $\lms$ from the following relation,
\be
F^\gamma_{exp} (x,Q^2) =  F^\gamma_{PQCD}(x,Q^2,\lms^2)+F^\gamma_{NP, ext}(x,Q^2,P^2=0),
\ee
where $F^\gamma_{NP, ext}(x,Q^2,P^2=0)$ is the extrapolated value of $F^\gamma_{NP}$ at $P^2=0$.

%%%%%%%%%%%%%%%%%%%%%%%%%%%%%%%%%%%%%%%%%%%%%%%%%%%%%%%%%%%%%%%%%%
\section{Model for the Non-perturbative Photon Structure Function}
%%%%%%%%%%%%%%%%%%%%%%%%%%%%%%%%%%%%%%%%%%%%%%%%%%%%%%%%%%%%%%%%%%

There are two methods to solve PSF. The first method is to solve the Altarelli-Parisi equations for the PSF numerically, 
while the second method uses both operator product expansion and the renormalization group equations
to calculate the moments of the PSF.
The second method is abstract but more rigorous, especially in high order calculations.
We will use the results from the second method for the perturbative part and
the vector meson dominance model for the non-perturbative part.
%The vector meson structure functions can be obtained from the quark distribution functions of the pion structure functions.

The steps to find $F^\gamma_2(x,Q^2,P^2)$ are following.
\begin{enumerate}[label=(\arabic*)]
\item Find the moments $F^\gamma_{PQCD}(n,Q^2,P^2)$ of the perturbative structure function using OPE and RGE.
\item Invert the moments using the inverse Mellin transform to obtain $F^\gamma_{PQCD}(x,Q^2,P^2)$.
\item Model $F^\gamma_{NP}(x,Q^2,P^2)$ using vector mesons, with a dominant contribution from the $\rho$ meson.
\item Finally, the total photon structure function $F^\gamma_2(x,Q^2,P^2)$ is obtained
by summing two contributions: the perturbative part $F^\gamma_{PQCD}(x,Q^2,P^2)$ and the non-perturbative part $F^\gamma_{NP}(x,Q^2,P^2)$.
\end{enumerate}

Now, we will explain how we perform step (3) in this section.
For a QCD process involving the kinematics with both high and low momentum,
it is necessary to separate the kinematic region into two parts.
The fundamental QCD scale parameter, $\lms$, acts as the reference point separating two energy regions.
This can be visualized in the $e\gamma$ scattering,
where a photon fluctuates into a quark-antiquark pair that undergoes complex gluon interactions.
The transverse momentum integration of partons over available phase space becomes \cite{BW:1987},
\be
\int_{Q^2_{min}}^{Q^2} {dp^2_T\over p^2_T+O(m^2_\gamma)+O(m^2_q)}
  = \int_{Q^2_{min}}^{\lms^2} {dp^2_T\over p^2_T+O(m^2_\gamma)+O(m^2_q)}
   + \int_{\lms^2}^{Q^2}{dp^2_T\over p^2_T+O(m^2_\gamma)+O(m^2_q)}.
\ee
The second term in the equation is perturbatively calculable while the first one is not.
Therefore, we separate the PSF into two parts:
a non-perturbative QCD part (NP) and a perturbative QCD part (PQCD).
rPSF and vPSF can be expressed respectively as
\ba
F^{\gamma}_{2}(x,Q^2) &=& F^{\gamma}_{2, PQCD}(x,Q^2)+F^{\gamma}_{2,NP}(x,Q^2),
\label{xrtot}\\
F_2^\gamma(x,Q^2,P^2) &=& F^\gamma_{2,PQCD}(x,Q^2,P^2)+ F^\gamma_{2,NP}(x,Q^2,P^2).
\label{xvtot}
\ea
How to identify $F^\gamma_{2,NP}$ is still an open question.
We can, of course, define $F^\gamma_{2,NP}$ from the above equations if we know
$F^\gamma_{2,PQCD}$ theoretically and $F_2^\gamma$ experimentally.

Now, let us investigate the vPSF in the two photons scattering.
The two photons scattering cross section can be written as the absorptive part of the two
photons forward scattering amplitude, shown in Fig. 2.
The amplitude contains a Feynman parametrization in the denominator.
Furthermore, the dispersion relation of the amplitude-more interestingly, the structure function $F_2^\gamma$-
possesses an analytic property featuring a branch cut.
We consider the two photon process with $q^2=-Q^2$ as the invariant mass$^2$
for the probe photon and $p^2=-P^2$ as the invariant mass$^2$ for the target photon.
For a fixed $Q^2$, the moment of the structure function is analytic in a cut $p^2$-plane.
The general form of the moments of the structure functions is \cite{Bjorken}
\be
F^\gamma_2(n,Q^2,P^2)=\int_0^1 dx x^{n-2} F^\gamma_2 (x,Q^2,P^2)
 =\int^\infty_0 d\sigma^2 {\rho (n,Q^2,P^2) \over P^2+\sigma^2},
\ee
where $\rho(n,Q^2,P^2)$ is the spectral density corresponding to the moment function $F^\gamma_2(n,Q^2,P^2)$.

The non-perturbative part of the structure function looks like \cite{Bjorken,BW:1987},
\ba
F^\gamma_{2,NP}(n,Q^2,P^2)&\equiv&\int_0^\infty {\rho_{NP} (n,Q^2,P^2) \over
P^2+\sigma^2} d\sigma^2.
\ea
At this point, we must remind the reader that $\rho_{NP}$ does not simply
stem from the purely hadronic contribution used in usual sense.
For example, the typical non-perturbative and the perturbative contributions represent
two extreme limits of the perturbation parameter.
However, we must integrate the above equation over the entire range of $\sigma$, rather than just over these two limits.

Although it is not rigorous, we have some information about the non-perturbative part of the spectral density, $\rho_{NP}$.
This $\rho_{NP}$ will be strongly suppressed above the masses of the vector mesons.
In general, we can have multi-$\rho$ pole contribution to $\rho_{NP}$-including those no
pole terms-many of which may come from perturbative calculations.
The double pole term ($\rho_{NP}\propto\delta^\prime (\sigma^2-m_\rho^2$) )
which arises from the two $\rho$ meson-pole contribution to the $\gamma\gamma$ forward
scattering amplitude, is expected to be the dominant one \cite{IW, Bjorken}.
Therefore, the general form of the non-perturbative part of vPSF becomes
\be
F^{\gamma}_{2,NP}(x,Q^2,P^2) = {F^{\gamma}_{2,NP}(x,Q^2) \ov (1+P^2/M^2)^2}.
\ee
The term $F^{\gamma}_{2,NP}(x,Q^2)$ represents the non-perturbative part of rPSF.
The non-perturbative rPSF, $F^{\gamma}_{2,NP}(x,Q^2)$, can be approximated by the contributions from light vector mesons.
Considering all light vector mesons with the same quantum numbers as the photon ($\rho$, $\omega$, $\phi$),
we can express the non-perturbative rPSF and the non-perturbative vPSF as an incoherent sum of
the light vector meson structure functions.
According to \cite{PWZ1,PWZ2,BW:1987}, the difference between using a coherent sum and an incoherent sum is not significant,
\ba
F^{\gamma}_{2,NP}(x,Q^2) 
&\simeq&\sum_V  {\alpha\pi\ov\gamma_V^2}F^V_2(x,Q^2),\\
F^{\gamma}_{2,NP}(x,Q^2,P^2) 
&\simeq&\sum_V  {\alpha\pi\ov\gamma_V^2}{F^V_2(x,Q^2) \ov (1+P^2/M_V^2)^2}.
\ea

\noindent {\bf VMD weights and quantitative $\rho$ dominance.} 
The coupling parameters $\gamma_V$ in the VMD weights can be fixed using the dilepton widths $V\to e^+e^-$.
A standard Lagrangian-level implementation of photon--vector-meson mixing is
\cite{BauerVMD,SakuraiVMD}
\be
\mathcal{L}_{\gamma V} \;=\; -\sum_{V=\rho,\omega,\phi}\frac{e\,m_V^2}{\gamma_V}\,V_\mu A^\mu,
\label{L_gV}
\ee
which is equivalent to the current--field identity
\be
J_{\rm em}^\mu \;=\; \sum_{V=\rho,\omega,\phi}\frac{m_V^2}{\gamma_V}\,V^\mu.
\label{current_field_identity}
\ee
From $V\to\gamma^*\to e^+e^-$, one obtains (Appendix)
\be
\Gamma(V\to e^+e^-)=\frac{4\pi\alpha^2}{3}\,\frac{m_V}{\gamma_V^2}.
\label{Gamma_ee}
\ee
Therefore, the weight factor can be rewritten as
\be
w_V\equiv \frac{\alpha\pi}{\gamma_V^2}
=\frac{3}{4}\,\frac{\Gamma(V\to e^+e^-)}{\alpha\,m_V},
\label{wV_from_Gee}
\ee
allowing a PDG-based estimate of relative weights.
The PDG values \cite{PDG_rho,PDG_omega,PDG_phi} are given as
\be
w_\rho : w_\omega : w_\phi \simeq 1 : 0.091 : 0.138.
\ee
In this work, we keep the dominant $\rho$ term as a baseline and treat the omission of $\omega,\phi$
as a modeling systematic; incorporating the full sum is straightforward within our framework.

\begin{table}[t]
\centering
\caption{
PDG-based VMD weights. We compute $\Gamma_{ee}=B_{ee}\Gamma_{\rm tot}$ and use
$w_V=\alpha\pi/\gamma_V^2=(3/4)\,\Gamma_{ee}/(\alpha m_V)$. The last column shows $w_V/w_\rho$
(normalized to $\rho$) \cite{PDG_rho,PDG_omega,PDG_phi}.
}
\smallskip
\begin{tabular}{lcccc}
\hline\hline
$V$ & $m_V$ [MeV] & $\Gamma_{\rm tot}$ [MeV] & $B_{ee}$ & $w_V/w_\rho$ \\
\hline
$\rho(770)$  & $\sim 775$ & $\sim 147.4$ & $4.72\times 10^{-5}$ & $1$ \\
$\omega(782)$& $782.66$ & $8.68$ & $7.38\times 10^{-5}$ & $0.091$ \\
$\phi(1020)$ & $1019.461$ & $4.249$ & $3.00\times 10^{-4}$ & $0.138$ \\
\hline\hline
\end{tabular}
\label{tab:vmd_weights}
\end{table}

Among the mesons, the $\rho$ meson contributes the most to $F^{\gamma}_{2,NP}$.
The  $\rho$ meson structure functions, $F^{\rho^0}_2$, can be obtained
from pion structure functions $F^{\pi^0}_2$ or $F^{\pi^\pm}_2$
using rotational and isospin symmetry.
\be
F^{\gamma}_{2,NP}(x,Q^2) \approx {\alpha\pi\ov\gamma_\rho^2}F^{\rho^0}_2(x,Q^2)
={\alpha\pi\ov\gamma_\rho^2}F^{\pi^+}_2(x,Q^2),
\ee
where we use $F^{\pi^+}_2$ for $F^{\rho^0}_2$ since $F^{\rho^0}_2=F^{\pi^0}_2=F^{\pi^+}_2$.

The contribution to the  $F^{\gamma}_{2,NP}$ can be expressed as the incoherent sum of the vector mesons \cite{PWZ1},
\be
F^{\gamma}_2(x,Q^2) \approx\left\{ {\alpha\pi\ov\gamma^2_\rho}+{\alpha\pi\ov\gamma^2_\omega}
+{2\ov5}{\alpha\pi\ov\gamma^2_\phi}\right\}F^{\rho_0}_2(x,Q^2) 
\ee
The value of $(\alpha\pi/\gamma^2_V)$ is approximately $(2.85\pm0.3)\times10^{-3}$, $(0.4\pm0.04)\times10^{-3}$,
and $(0.6\pm0.05)\times10^{-3}$ for $\rho$, $\omega$, and $\phi$ mesons respectively \cite{BW:1987}.
Including all vector mesons raises the total contribution to approximately 20\%.
%$M_\rho=768.1\pm0.5$ MeV, M_\omega=781.95\pm0.14$ MeV, and M_\phi=1019.413\pm0.008$ MeV

Considering only the dominant $\rho$-meson contribution, we have
\be
F^{\gamma}_{2,NP}(x,Q^2,P^2)\approx
{\alpha\pi\ov\gamma_\rho^2}{F^{\pi^+}_2(x,Q^2)\ov(1+P^2/M_\rho^2)^2}.
\ee
$F^{\pi^+}_2$ is derived from the proton structure function \cite{NSZ,Arash} and expressed as
\be
 F^{\pi^+}_2(x,Q^2)\approx {2\ov3}F^P_2({2\ov3}x,Q^2).
\ee
We use  the proton structure function parametrized by the H1 collaboration \cite{H1},
\be
F^P_2(x,Q^2)=[ax^b+cx^d(1+e\sqrt{x})(\log Q^2+f\log^2Q^2+h/Q^2](1-x)^g.
\label{F2H1}
\ee
The values of parameters in Eq.  (\ref {F2H1}) are given in Table 1.
\begin{table}
\centering
\caption{The parameter values of the proton structure function in Eq. (\ref{F2H1})}
\smallskip
\begin{tabular}{cccccccc}
\hline\hline
  a&b&c&d&e&f&g&h \\
\hline
~3.10~&~0.76~&~0.124~&~$-$0.188~&~$-$2.91~&~$-$0.043~&~3.69~&~1.40~ \\
\hline\hline
\end{tabular}
\label{table1}
\end{table}
The final expressions for the total vPSF becomes
\be
F^{\gamma}_2(x,Q^2,P^2) \approx F^{\gamma}_{2,PQCD}(x,Q^2,P^2)
+{2\ov3}{\alpha\pi\ov\gamma_\rho^2}{F^P_2({2\ov3}x,Q^2)\ov(1+P^2/M_\rho^2)^2}.
\ee

%%%%%%%%%%%%%%%%%%%%%%%%%%%%%%%%%%%%%%%%%%%%%%%%%%%%%%%%%%%%%%%%%%
\section{Numerical Analysis for the Extraction of $\Lambda_{\overline{\textrm{MS}}}$}
%%%%%%%%%%%%%%%%%%%%%%%%%%%%%%%%%%%%%%%%%%%%%%%%%%%%%%%%%%%%%%%%%%

The rigorous method for calculating the higher order contributions to the PSF is to use the OPE and RGE.
The structure function obtained this way depends on the moments $n$ rather than the Bjorken variable $x$.
Therefore, to obtain the structure function in terms of $x$,
we must invert the moments to recover the usual structure function of $x$.
This inversion of the moment functions can be performed numerically.
While there are several ways for numerical inversion, as discussed in \cite{Kim,BJK},
here we use the $\chi^2$ Monte Carlo method, as in \cite{BJK}, to minimize $\chi^2$
between the moments calculated from OPE/RGE and those derived from the fitting functions.

The common form of the structure functions used most by physicists is
\be
F(x)=x(1-x)^{-\alpha}(\sum_{k=0}^N a_k x^{k-\beta})
\label{str_ft}
\ee
with $|\alpha|<1$, $\beta>-1$, and positive integer $N$.
The moment of this function is
\be
f(n)=\sum_{k=0}^N a_k {\Gamma(1-\alpha)\Gamma(n+k-\beta) \ov \Gamma(n+k+1-\alpha-\beta)}
\quad{\rm with}\quad (n+k>\beta+1).
\label{momft}
\ee
There are several unknowns which can be determined by using
the least $\chi^2$ fitting method \cite{BJK,RS,Pressetal}.
$\chi^2$ can be expressed in terms of fitted moment functions with the moment data,
\be
\chi^2=\sum_{j=0}^M {[f(j)-f_j]^2\ov M-N}.
\label{chi2}
\ee
Where $M$ is the number of data and $N$ is the number of unknown constants ($N<M$).
$f_j$ is the datum of the $j$-th moment given in Eq. (\ref{momft}).
We are not going into the the details of this method which can be found in  \cite{BJK}.

\begin{figure}[h]                             
%	\centering
\begin{center}
	\includegraphics[width=0.7\textwidth]{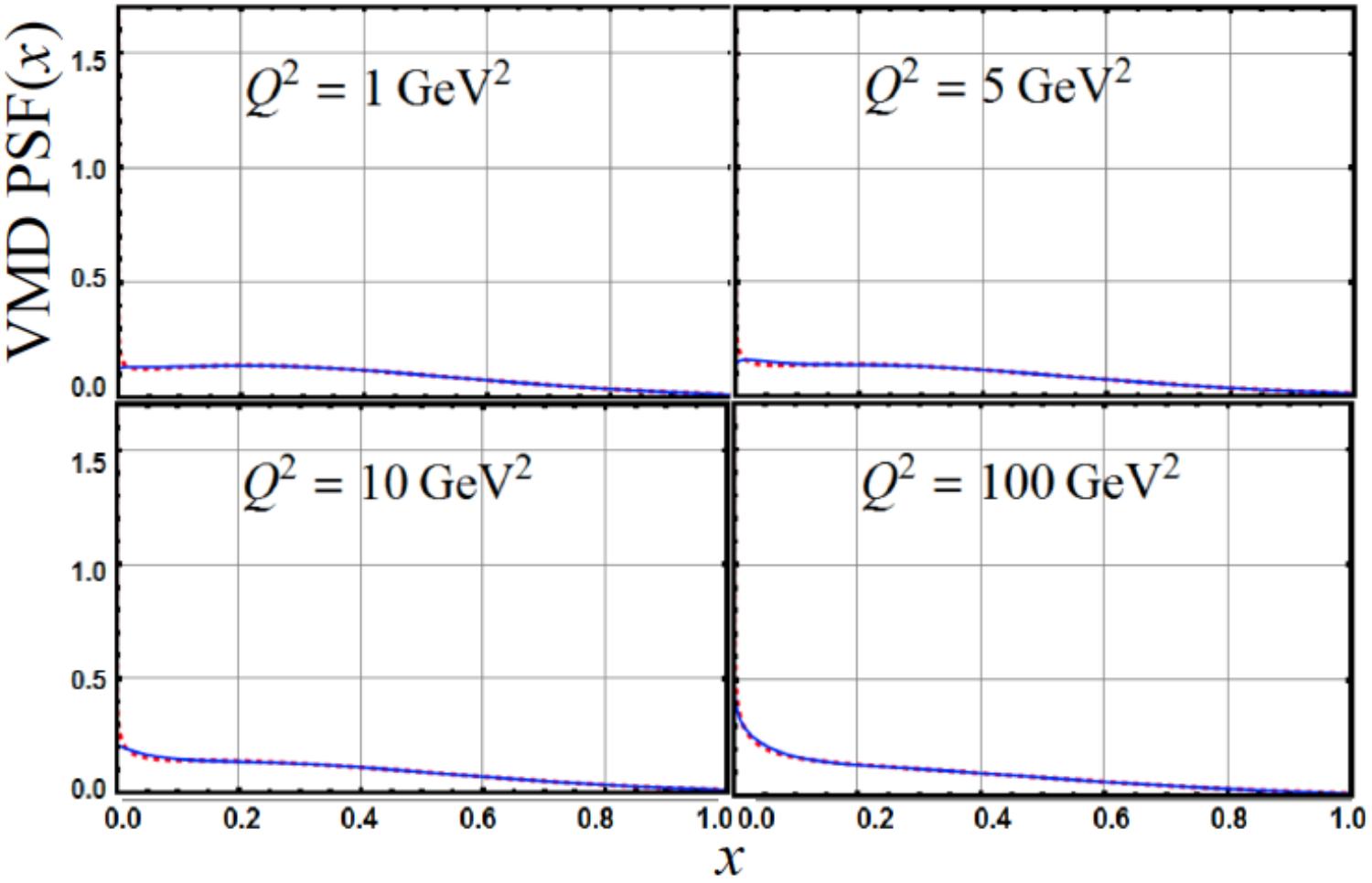}
	%\hspace*{-2cm}
\end{center}
	\caption{$F_2,_{NP}(x,Q^2)$ based on VMD for a few values of $Q^2$}
	\label{fig_f2np}	
\end{figure}

Fig. \ref{fig_f2np} shows the non-perturbative results from our calculation.
The structure functions of the VMD parts of the rPSF ranges from 1~GeV$^2$ to 100~GeV$^2$.
The solid and dashed lines represent inverted moment functions and original moment functions, respectively.
They are almost overlapping, making them indistinguishable from one another, except in the low-$x$ region
where uncertainty is greatest.

\begin{figure}[h]                             
%	\centering
\begin{center}
	\includegraphics[width=0.7\textwidth]{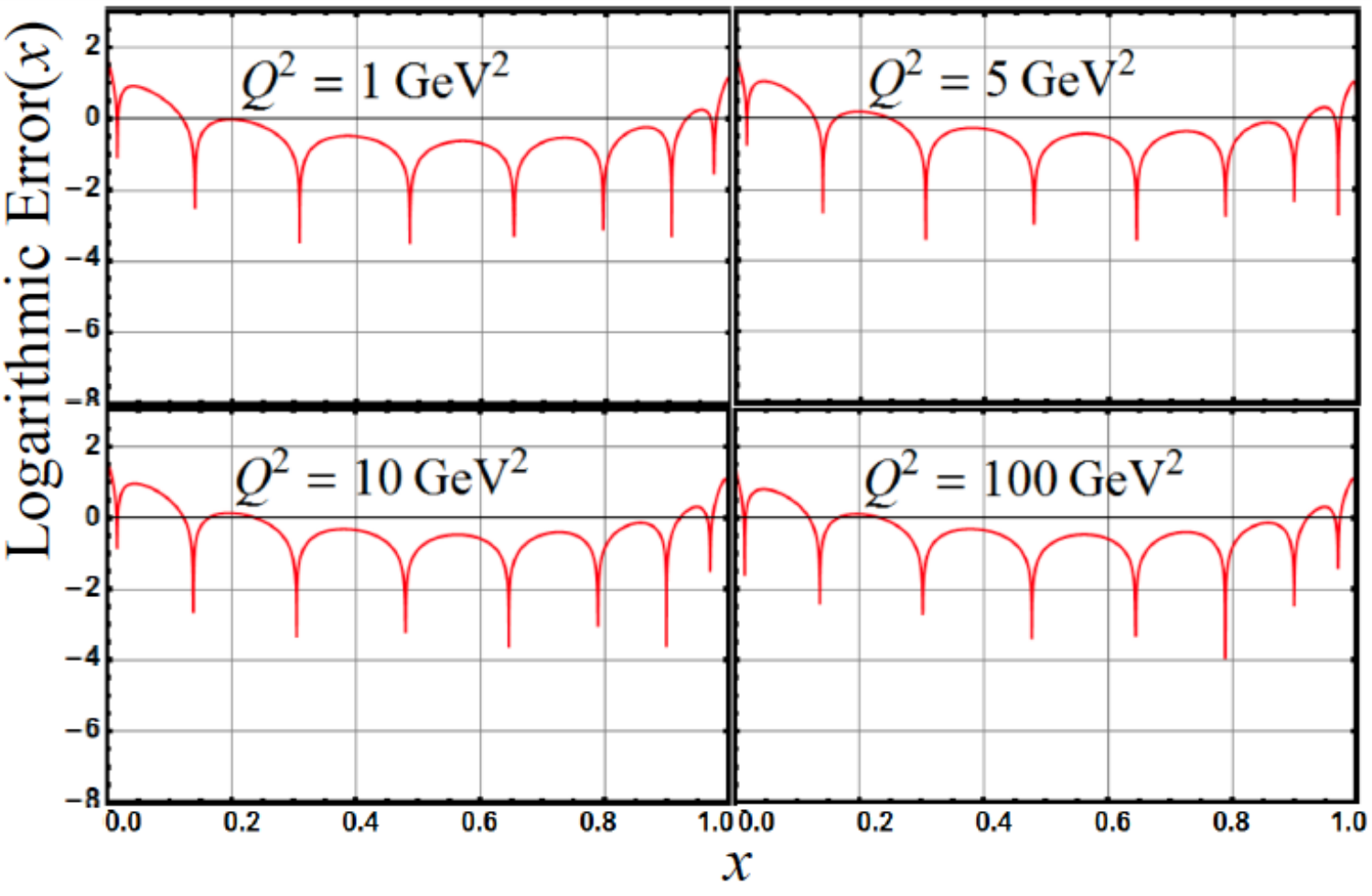}
	%\hspace*{-2cm}
\end{center}
	\caption{Logarithmic Error of $F_2,_{NP}(x,Q^2)$}
	\label{fig_f2nperr}	
\end{figure}

Fig. \ref{fig_f2nperr} shows the inversion error for the $\chi^2$ method.
It shows the corresponding logarithmic errors between the original
and inverted structure functions of the VMD part of the rPSF
in the region from 1~GeV$^2$ to 100~GeV$^2$.
We can see that the logarithmic error is satisfactory in the most of the $x$-region except for the regions near $x=0$ and $x=1$.
The error near $x=1$ is not a problem since the structure function is almost zero,
the uncertainty near $x=0$ is not a big concern either as the structure function is finite.\\

%========================================
%========================================

%========================================
%========================================

\begin{figure}[h]                             
%	\centering
\begin{center}
	\includegraphics[width=0.7\textwidth]{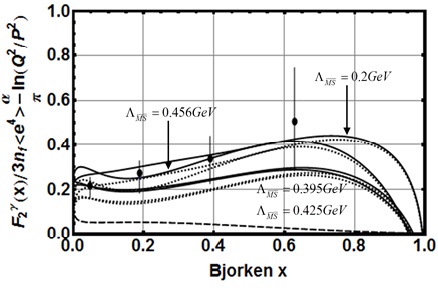}
	%\hspace*{-2cm}
\end{center}
	\caption{The total vPSF for $Q^2=5$ GeV$^2$ and $P^2=0.35$ GeV$^2$ compared to PLUTO results.}
%We omitted the L3 data \cite{L3} to show a simple example of the total vPSF.}
	\label{fig_vpsfl_tot}	
\end{figure}

Now, the perturbative QCD part of the vPSF is obtained when we calculate the inversion of moments in the next-to-leading order.
The non-perturbative part of the vPSF is also obtained from VMD including only $\rho$ meson contribution.
Then, we can obtain the total vPSF by adding these two contributions.
Fig. \ref{fig_vpsfl_tot} shows the total vPSF for four values of $\lms$ for
$Q^2=5$ GeV$^2$, $P^2=0.35$ GeV$^2$ compared to the data from PLUTO \cite{PLUTO}.
The plot with solid, dotted, dashed lines indicate the total vPSF $F^\gamma_2$,
the perturbative part of vPSF, $F^\gamma_{2, PQCD}$,
and the non-perturbative part of vPSF, $F^\gamma_{2, NP}$, respectively.
They fit reasonably well with the experimental data especially for two cases of $\lms=0.2$ GeV and $\lms=0.456$ GeV.

%========================================
%%========================================

%========================================
\begin{table}
\centering
\caption{%
QCD scale parameters and the Strong coupling constants 
from the least $\chi^2$ method for the a few $Q^2$ values.}
\smallskip
\begin{tabular}{cccccccc}
\hline\hline
  $Q^2$&$\quad\Lambda_{\overline{\textrm{MS}}}
  \quad$&$\alpha_S(Q^2,\Lambda_{\overline{\textrm{MS}}})$&$\chi^2$ \\
\hline
1&0.4058&0.634&1.45647 \\
5&0.3450&0.298&5.20817\\
10&0.3792&0.266&9.92634\\
90&0.3435&0.179&9.35728\\
100&0.3833&0.182&3.18593\\
1000&0.4064&0.141&6.20691\\
%1&6.2160&0.4058&0.634&1.45647 \\
%5&5.7591&0.3450&0.298&5.20817\\
%10&3.8520&0.3792&0.266&9.92634\\
%20&3.2310&0.3120&0.217&2.04026\\
%30&4.4929&0.3954&0.220&1.43721\\
%40&6.2064&0.3345&0.199&1.64371\\
%50&5.5577&0.3569&0.197&1.25501\\
%60&6.4416&0.3345&0.188&3.08062\\
%70&5.5375&0.3735&0.190&1.06847\\
%80&4.0203&0.4086&0.191&2.48657\\
%90&5.2105&0.3435&0.179&9.35728\\
%100&4.7335&0.3833&0.182&3.18593\\
%200&4.8608&0.3554&0.164&2.25577\\
%500&3.9977&0.3342&0.146&1.57768\\
%1000&5.2381&0.4064&0.141&6.20691\\
%10000&5.3440&0.3736&0.113&5.90991\\
\hline\hline
\end{tabular}
\label{table2}
\end{table}

While the QCD scale parameter ($\lms$) has a rather wide range of values around $\Lambda_{\overline{\textrm{MS}}}=200-400~\textrm{MeV}$, and 
multiple methods exist for determining it-such as theoretical calculations and experimental data fit \cite{lambda-val1,lambda-val2,lambda-val3,lambda-val4,lambda-val5}-no universally accepted standard value has been established. In this work, however, we employ a least-$\chi^2$ fit to the $Q^2$–dependence of our observable within the range $1~\mathrm{GeV}^2 \le Q^2 \le 10^4~\mathrm{GeV}^2$,
we obtain an overall value 
\be
\Lambda_{\overline{\textrm{MS}}}=365.1^{+43.5  {\rm MeV}}_{-53.1 {\rm MeV}}.\\
\label{obt_lam}
\ee
The individual values of $\Lambda_{\overline{\mathrm{MS}}}$ extracted at fixed $Q^2$
fluctuate moderately around this central value and exhibit no clear systematic trend.
While Table~\ref{table2} displays only a representative subset,
these values were calculated over a broad range from $Q^2=1$ GeV$^2$ to $Q^2=10000$ GeV$^2$.

The scale parameter $\lms$ depends on the number of active quark flavors $n_f$.
The value of $Q^2$ determine the appropriate $n_f$;
for example, assuming a charm quark mass of 1.5 GeV, the choice $n_f=4$ is reasonable for $Q^2 >$ 9 GeV$^2$ 
to account for the quark pair production.
Conversely, for $Q^2 < 9$ GeV$^2$, it is appropriate to set $n_f=3$.
We exclude the region of $Q^2 < 1$ GeV$^2$ because the condition $\lms^2 \ll P^2 \ll Q^2$ is difficult to satisfy.
Therefore, $\lms$ obtained in this work corresponds to $n_f=3,4,5$ depending on the $Q^2$ value.
The final result is the average of all $\lms$ values.

Evolving the coupling to the $Z$-boson mass then yields 
\be
\alpha_S(M_Z,\Lambda_{\overline{\textrm{MS}}})=0.1146^{+0.0021}_{-0.0028},
\label{obt_coup}
\ee
which is statistically compatible with the value of the Particle Data Group (PDG) 2023
average in Eq. (9.25) of Ref.~\cite{PDG2024QCD} (also in \cite{PDG2024, AKS,T-etal}), \be
\alpha^{Ref}_S(M_Z)=0.1180 \pm 0.0009.
\label{alpha_ref}
\ee
This is because the difference between the central values, which is
$\Delta \alpha_S = 0.1180 - 0.1146 = 0.0034$, and the combined error is given by
\[
\sigma_{\text{tot}}^{(+)} = \sqrt{(0.0021)^2 + (0.0009)^2} \approx 0.0023,
\qquad
\sigma_{\text{tot}}^{(-)} = \sqrt{(0.0028)^2 + (0.0009)^2} \approx 0.0029,
\]
leading to the significance of the difference
\[
\frac{\Delta \alpha_S}{\sigma_{\text{tot}}^{(+)}} \approx 1.5,
\qquad
\frac{\Delta \alpha_S}{\sigma_{\text{tot}}^{(-)}} \approx 1.2.
\]
Thus, depending on whether the upper or lower combined uncertainty is used, the shift corresponds to about $1.2\sigma\sim1.5\sigma$,
which is well below the conventional $2\sigma$ ($\sim 95\%$ confidence level) threshold for a statistically significant discrepancy.
This indicates that our determination of $\alpha_S(M_Z,\Lambda_{\overline{\textrm{MS}}})$ in Eq.~\eqref{obt_coup} evaluated at the extracted QCD scale parameter in Eq.~\eqref{obt_lam},
which lies within the commonly quoted range of $200\sim400$ MeV, is statistically consistent with the PDG average in Eq. \eqref{alpha_ref}.
This supports the conclusion that the proposed method for extracting $\Lambda_{\overline{\mathrm{MS}}}$ is reliable at the present level of precision and provides a promising basis for more refined future studies.

%%%%%%%%%%%%%%%%%%%%%%%%%%%%%%%%%%%%%%%%%%%%%%%%%%%%%%%%%%%%%%%%%%
\section{Conclusion}
%%%%%%%%%%%%%%%%%%%%%%%%%%%%%%%%%%%%%%%%%%%%%%%%%%%%%%%%%%%%%%%%%%

The total photon structure function (PSF), $F^\gamma_2(x,Q^2,P^2)$, can be decomposed into a perturbative part (PQCD), $F^\gamma_{2,PQCD}(x,Q^2,P^2)$,
and a non-perturbative part (NP), $F^\gamma_{2,NP}(x,Q^2,P^2)$.
The perturbative component of the PSF in the next-to-leading order, $F^\gamma_{2,PQCD}(x,Q^2,P^2)$, can be found in \cite{BB1,BB2,UW}.
For the non-perturbative part, we propose a model based on the vecor meson dominance
as the vector meson sharing the same quantum numbers as the photon.
We use the $\rho$ meson structure function to represent $F^\gamma_{2,NP}(x,Q^2,P^2)$.
We then employ a $\chi^2$ minimization fitting procedure for
the moments of the real and virtual photon structure functions
to obtain the corresponding structure functions in terms of Bjorken $x$.
By adding these two contributions, we obtain the total PSF.

For the kinematic range of $\lms^2 \ll P^2 \ll Q^2$, the experimental value of the
structure function is approximately equal to its perturbative part of the structure function,
$F^\gamma_{2,exp} \approx F^\gamma_{2,PQCD}$, due to the weak $\lms^2$ dependence in $F^\gamma_{2,PQCD}$.
Additionally, $F^\gamma_{2,NP}$ is a slowly varying function of
$P^2$ within the range of 0.5 GeV$^2$ to 2 GeV$^2$ \cite{PWZ1,PWZ2,BW:1987}.
Consequently, we can extrapolate the value of $F^\gamma_{2,NP}$ at $P^2=0$.
If the rPSF, $F^\gamma_{2,NP, exp}$, is measured at $Q^2$,
then $\lms$ can be extracted using the following relation \cite{Kim},
\be
F^\gamma_{2,exp} (x,Q^2) = F^\gamma_{2,PQCD}(x,Q^2,\lms^2) + F^\gamma_{2,NP, ext}(x,Q^2,P^2=0),
\ee
where $F^\gamma_{2,NP, ext}$ stands for the extrapolated value of $F^\gamma_{2,NP}$ at $P^2=0$.

Using experimental result, we can determine a consistent value for $\lms$.
Specifically, a measurement of $F^\gamma_{2,exp} (x,Q^2)$ at a given $Q^2$ yields a value for $\lms$,
while subsequent measurements at a different $Q^2$ scales provide a robust consistent check.
Therefore, we recommend a systematic series of experiments to measure
photon structure functions for various values of $Q^2$ and $P^2$.

Finding the QCD scale parameter is still an open question.
The objective of this study is to confirm the initial proposal \cite{IW}
regarding the extraction of $\lms$ from photon structure functions, rather than to obtain a precise value.
Nonetheless, the value obtained using our method is reasonably good.
Future analyses, however, will need to be more fine-tuned.
We hope that a more complete analysis-such as including all the vector mesons-will yield better results.
We also hope that more experiments on photon structure functions will be performed for various values of $Q^2$ and $P^2$,
which will shed more light on a better value for the QCD scale parameter, $\Lambda_{\overline{\textrm{MS}}}$.

\subsection*{Acknowledgments} 
H.J. is supported by the National Research Foundation of Korea (NRF) through the Grants: RS-2020-NR049598 (CQUeST, Sogang University), RS-2023-NR077094, and RS-2024-00441954.

%%%%%%%%%%%%%%%%%%%%%%%%%%%%%%%%%%%%%%%%%%%%%%%%%%%%%%%%%%%%%%%%%%
\section{Appendix: Definition of $\gamma_V$ and derivation of $\Gamma(V\to e^+e^-)$}
%%%%%%%%%%%%%%%%%%%%%%%%%%%%%%%%%%%%%%%%%%%%%%%%%%%%%%%%%%%%%%%%%%
\label{app:gammaV}

In this appendix, we derive the relation between the VMD coupling parameter $\gamma_V$ and the dilepton width $\Gamma(V\to e^+e^-)$ used in Sec.~3.
In VMD, the electromagnetic current is represented by the light neutral vector mesons \cite{BauerVMD,SakuraiVMD}:
\be
J_{\rm em}^\mu = \sum_{V=\rho,\omega,\phi} \frac{m_V^2}{\gamma_V} V^\mu,
\ee
so that the photon interaction becomes
\be
\mathcal{L}_{\rm int} = -e A_\mu J_{\rm em}^\mu
= - \sum_V \frac{e m_V^2}{\gamma_V} V_\mu A^\mu.
\ee
The photon couples to leptons as
\be
\mathcal{L}_{\gamma \ell\ell} = - e A_\mu\, \bar{\ell}\gamma^\mu \ell .
\ee
The decay $V\to e^+e^-$ proceeds via $V\to\gamma^*\to e^+e^-$.
At tree level, combining the $V$--$\gamma$ mixing vertex with the photon propagator at $q^2=m_V^2$
yields an effective $V$--lepton coupling proportional to `$(e/\gamma_V)e$', and
the resulting amplitude becomes
\be
i\mathcal{M}(V\to e^+e^-)
= \left(\frac{e}{\gamma_V}\right) e\,
\varepsilon_\mu(q)\,\bar{u}(p_-)\gamma^\mu v(p_+).
\ee
Squaring this, averaging over the three initial polarizations and summing over final spins,
one obtains the standard width (neglecting $m_e$)
\be
\Gamma(V\to e^+e^-)=\frac{4\pi\alpha^2}{3}\,\frac{m_V}{\gamma_V^2}.
\ee
This implies
\be
w_V\equiv \frac{\alpha\pi}{\gamma_V^2}=\frac{3}{4}\,\frac{\Gamma(V\to e^+e^-)}{\alpha\,m_V},
\ee
which is used in Sec.~3 to estimate the relative weights of $\rho,\omega,\phi$ from PDG data.

%%%%%%%%%%%%%%%%%%%%%%%%%%%%%%%%%%%%%%%%%%%%%%%%%%%%%%%%%%%%%%%%%%
 %===============

\end{document}